\documentclass[review]{elsarticle}

\usepackage{nohyperref,
amsmath,amssymb,xcolor,bm,soul}    
	\soulregister\cite7
    \soulregister\ref7
    \soulregister\eqref7
    \soulregister\gls7
    \soulregister\glsentrytext7
    \soulregister\pageref7

\usepackage[acronyms,toc,indexonlyfirst]{glossaries}
\newglossary[nlg]{notation}{not}{ntn}{Notation}
\makenoidxglossaries

\newglossaryentry{aa}
	{
		name={amino acid},
		description={}
	}

\newglossaryentry{al}
	{
		name={artificial life},
		plural={artificial lives},
		description={a field of study wherein researchers examine systems related to natural life, its processes, and its evolution, through the use of simulations with computer models, robotics, and biochemistry}
	}
	
\newglossaryentry{amm}
	{
		name={ammonia},
		description={}
	}	
	
\newglossaryentry{asn}
	{
		name={asparagine},
		description={}
	}

\newglossaryentry{asp}
	{
		name={aspartate},
		description={}
	}		

\newglossaryentry{at}
	{
		name={advice taking},
		description={}
	}	

\newglossaryentry{beta}
	{
		name={degree of social influence},
		description={}
	}

\newglossaryentry{cl}
	{
		name={closed-loop},
		description={}
	}	

\newglossaryentry{cu}
	{
		name={copper},
		description={}
	}	
	
\newglossaryentry{dm}
	{
		name={decision-making},
		description={}
	}

\newglossaryentry{dr}
	{
		name={distributionally robust},
		description={}
	}
	
\newglossaryentry{drd}
	{
		name={distributionally robust design},
		description={}
	}
	
\newglossaryentry{ds}
	{
		name={dynamic system},
		description={}
	}		

\newglossaryentry{ead}
	{
		name={egocentric advice discounting},
		description={}
	}

\newglossaryentry{fg}
	{
		name={``Fitness Game''},
		description={}
	}

\newglossaryentry{glc}
	{
		name={glucose},
		description={}
	}
	
\newglossaryentry{gln}
	{
		name={glutamine},
		description={}
	}	
	
\newglossaryentry{glu}
	{
		name={glutamate},
		description={}
	}

\newglossaryentry{gly}
	{
		name={glycosylation},
		description={}
	}	
	
\newglossaryentry{id}
	{
		name={system identification},
		description={}
	}

\newglossaryentry{iv}
	{
		name={\emph{in vivo}},		
		description={}
	}
	
\newglossaryentry{lac}
	{
		name={lactate},
		description={}
	}	
	
\newglossaryentry{mm}
	{
		name={Michaelis--Menten},
		description={}
	}

\newglossaryentry{ms}
	{
		name={multiscale},
		description={}
	}

\newglossaryentry{mvt}
	{
		name={mean value theorem},
		description={}
	}		
	
\newglossaryentry{ne}
	{
		name={Nash equilibrium},
		plural={Nash equilibria},
		description={}
	}
	
\newglossaryentry{ol}
	{
		name={open-loop},
		description={}
	}
\newglossaryentry{oln}
	{
		name={open loop},
		description={}
	}	

\newglossaryentry{or}
	{
		name={optimally robust},
		description={}
	}
	
\newglossaryentry{ord}
	{
		name={optimally robust design},
		description={}
	}

\newglossaryentry{sfk}
	{
		name={soft feedback},
		description={}
	}

\newglossaryentry{SFK}
	{
		name={Soft Feedback},
		description={}
	}

\newglossaryentry{sr}
	{
		name={soft regulation},
		description={}
	}
	
\newglossaryentry{ss}
	{
		name={steady-state},
		description={}
	}	
		
\newglossaryentry{vp}
	{
		name={\emph{vox populi}},
		description={}
	}	

\newglossaryentry{woc}
	{
		name={wisdom of crowds},
		description={}
	}
	
\newglossaryentry{bao}
	{
		name={Prof.~Babatunde Ogunnaike},
		description={}
	}	

\newglossaryentry{bb}
	{
		name={Prof.~Bhavik Bakshi},
		description={}
	}

\newglossaryentry{gi}
	{
		name={Prof.~Garud Iyengar},
		description={}
	}	
	
\newglossaryentry{khl}
	{
		name={Prof.~Kelvin Lee},
		description={}
	}	
	
\newglossaryentry{ubb}
	{
		name={Upper Big Branch Mine},
		description={}
	}

\newglossaryentry{vv}
	{
		name={Prof.~Venkat Venkatasubramanian},
		description={}
	}

\newacronym{abm}{ABM}{agent-based model}
\newacronym{amt}{AMT}{Amazon Mechanical Turk}

\newacronym{cho}{CHO}{Chinese hamster ovary}
\newacronym{cqa}{CQA}{critical quality attribute}

\newacronym{doe}{DoE}{design of experiments}

\newacronym{er}{ER}{endoplasmic reticulum}

\newacronym{iid}{IID}{independent and identically distributed}

\newacronym{jas}{JAS}{judge-advisor system}

\newacronym{kkt}{KKT}{Karush--Kuhn--Tucker}

\newacronym[plural=monoclonal antibodies,firstplural=monoclonal antibodies (mAbs)]{mab}{mAb}{monoclonal antibody}
\newacronym{mc}{MC}{Monte Carlo}
\newacronym{mse}{MSE}{mean squared error}
\newacronym{msha}{MSHA}{Mine Safety and Health Administration}

\newacronym{nbs}{NBS}{Nash bargaining solution}
\newacronym{niosh}{NIOSH}{National Institute for Occupational Safety and Health}
\newacronym{nsf}{NSF}{National Science Foundation}

\newacronym{ode}{ODE}{ordinary differential equation}

\newacronym{pde}{PDE}{partial differential equation}
\newacronym{pfr}{PFR}{plug flow reactor}
\newacronym{pha}{PHA}{process hazard analysis}
\newacronym{pse}{PSE}{process systems engineering}
\newacronym{pso}{PSO}{particle swarm optimization}

\newacronym{tcd}{TCD}{total cell density}

\newacronym{vcd}{VCD}{viable cell density}

\newglossaryentry{state}
  {
    type=notation,
	name={\ensuremath{x}},
    description={state},
    sort={x}
  }
  
\newglossaryentry{output}
  {
    type=notation,
	name={\ensuremath{y}},
    description={output},
    sort={y}
  }  

\newglossaryentry{input}
  {
    type=notation,
	name={\ensuremath{u}},
    description={input},
    sort={u}
  }

\newglossaryentry{dis}
  {
    type=notation,
	name={\ensuremath{\omega}},
    description={disturbance},
    sort={omega}
  }

\newglossaryentry{noise}
  {
    type=notation,
	name={\ensuremath{\nu}},
    description={noise},
    sort={nu}
  }  

\journal{Physica A: Statistical Mechanics and its Applications}
\begin{document}

\begin{frontmatter}

\title{How much income inequality is fair?\\Nash bargaining solution and\\its connection to entropy}

\author[cu]{Venkat Venkatasubramanian\corref{cor}}\cortext[cor]{Corresponding author} 
\ead{venkat@columbia.edu}
\author[udel]{Yu Luo}
\ead{yuluo@udel.edu}
\address[cu]{Department of Chemical Engineering, Columbia University, New York, NY 10027} 
\address[udel]{Department of Chemical and Biomolecular Engineering, University of Delaware, Newark, DE 19716}

\begin{abstract}
The question about fair income inequality has been an important open question in economics and in political philosophy for over two centuries with only qualitative answers such as the ones suggested by Rawls, Nozick, and Dworkin. We provided a quantitative answer recently, for an ideal free-market society, by developing a game-theoretic framework that proved that the ideal inequality is a lognormal distribution of income at equilibrium. In this paper, we develop another approach, using the \gls{nbs} framework, which also leads to the same conclusion. Even though the conclusion is the same, the new approach, however, reveals the true nature of \gls{nbs}, which has been of considerable interest for several decades. Economists have wondered about the economic meaning or purpose of the \gls{nbs}. While some have alluded to its fairness property, we show more conclusively that it is all about fairness. Since the essence of entropy is also fairness, we see an interesting connection between the Nash product and entropy for a large population of rational economic agents.

\end{abstract}

\glsresetall

\end{frontmatter}

\newcommand{\rep}[2]{\textcolor{red}{\st{#1}}\hl{#2}}

\section{Introduction}\label{sec_back}

	Extreme economic inequality is widely seen as a serious concern to the future of stable and vibrant capitalist democracies. In 2015, the World Economic Forum in Davos identified deepening income inequality as the number one challenge of our time.  As many political observers remarked, the social and political consequences of extreme economic inequality  and the uneven sharing of prosperity seem to have played a role in the outcome of the U.S. presidential election in 2016.  Many in the U.S. feel that the nation's current level of economic inequality is unfair and that capitalism is not working for 90\% of the population~\cite{piketty2014capital,stiglitz2015great,reich2015saving}.

	Yet some income inequality is inevitable, even desirable and necessary for a successful capitalist  society. As different people have different talents and skills, and different capacities for work, they make different contributions in a society, some more, others less. Therefore, it is only fair that those who contribute more earn more. \emph{But how much more?} In other words, what is the fairest inequality of income? This critical question is at the heart of the inequality debate. The debate is not so much about inequality \emph{per se} as it is about fairness.

	Consider a simple example to illustrate this point. John is hired as a temporary worker to perform a job for one hour and makes \$100. Lilly also performs the same job, at the same level of quality, but works for two hours and makes \$200. Is there inequality in their incomes? Of course, there is. But is the inequality \emph{fair}? Of course, it is. Lilly earned more because she contributed more. Their incomes are \emph{not equal}, but \emph{equitable}. They are both paid at the same rate per hour of work, which is the basis for fairness here. 
    
    In this simple case, it was easy to ensure equity, but how do we accomplish this, in general, in a free market society consisting of millions of workers of varying degrees of talent, skill, and capacity for work? Is there a measure of fairness that can guide us to accomplish this? Is there an income distribution that ensures equity? Given the complexity of the problem, one might anticipate the answer to be no for both questions. But, surprisingly, the answer is yes, under certain conditions. 
    
    The first author has shown, in a series of papers \cite{venkatasubramanian2009fair,venkatasubramanian2010fairness,venkatasubramanian2015much} and a book \cite{venkatasubramanian2017inequality}, that the measure of fairness we are seeking is entropy and the equitable income distribution is lognormal at equilibrium in an ideal free market. These results were arrived at by analyzing the problem in two different, but related, perspectives: statistical mechanics and potential game theory. 
    
    In this paper, we demonstrate yet another approach, namely, the \gls{nbs} formalism~\cite{nash1950bargaining,muthoo1999bargaining}. We consider this paper to be valuable in two respects. One, this is the first time one has proposed the \gls{nbs}  formalism for the income distribution problem. Even though the \gls{nbs} formalism has been well-known for nearly seventy years, and has been used extensively in many fair allocation problems (see, for example, \cite{muthoo1999bargaining}), we find it quite surprising that it has not been used to address the central question of fair distribution of income in a free-market economy. Thus, we consider our \gls{nbs} formulation to this problem as an important contribution to economic literature. Secondly, while the final result of our analysis in itself is not new, this new approach, however, reveals something unexpected, namely, the true meaning of the Nash product and its connection with entropy. 
    
    In the \gls{nbs} formalism, one arrives at the solution by maximizing the \emph{product} of utilities, known as the \emph{Nash product} (more on this in Section~\ref{sec_nbs}). Over the years, economists have wondered about the true meaning, i.e.,~about the economic content, of the Nash product. The \emph{sum} of utilities of different agents makes economic sense as it gives us the total utility of the system, but why a \emph{product} of utilities? What does it stand for?        
    
    As Trockel~\cite{trockel2003meaning} observed: 
    \begin{quote}While all other characterizations, via axioms or via support by equilibria of non-cooperative games appear to reflect different aspects of this solution and to open new ways of interpretation, its most simple description via the Nash product seems to have escaped up to now a meaningful interpretation \ldots Yet, concerning its direct interpretation the situation is best described by the quotation of Osborne and
Rubinstein~\citep{osborne1994gametheory} (1994, p. 303): ``Although the maximization of a product of utilities is a simple mathematical operation it lacks a straightforward interpretation; we view it simply as a technical device.''
\end{quote}
Most people use it as a convenient mathematical device, but what is it really?

	Interestingly, the meaning of potential ($P^{\ast}$) in game theory also had posed a similar puzzle as Monderer and Shapley~\cite{monderer1996potential} had pointed out: 
	\begin{quote}This raises the natural question
about the economic content (or interpretation) of $P^{\ast}$: What do the firms try to jointly maximize? We do not have an answer
to this question. 
\end{quote}
This is again related to a similar fundamental question raised by Samuelson~\cite{samuelson1972maximum} decades ago in his Nobel lecture: 
\begin{quote}
What it is that Adam Smith's ``invisible hand'' is supposed to be maximizing?
\end{quote}
    
  We showed in our earlier work~\cite{venkatasubramanian2015much,venkatasubramanian2017inequality} that Samuelson, and Monderer and Shapley, were right in suspecting that something quite interesting and deep was missing in our understanding of these economic theories---what was missing  was the understanding of how the concept of fairness was intimately connected with all this. We showed that  both potential and entropy stand for the concept of fairness in a distribution and that  this is  what the "invisible hand" is maximizing. 
  
  Building on this insight, we offer, in this paper, a novel interpretation of the Nash product and its connection with entropy and fairness. We show how both employ the same mathematical device to accomplish the same objective. Our earlier work showed the deep connection between potential game and statistical mechanics via the concept of entropy. In this paper, we show how these two frameworks are deeply connected with the \gls{nbs} framework via the connection between entropy and the Nash product. Thus, all three puzzles---the ``invisible hand,'' the potential, and the Nash product---are related in a deep and interesting manner to each other, and to entropy, and all these are all related to the same critical economic concept, namely, fairness. 

\section{Potential Game Theoretic Framework: Summary of Past Work}

	Before we can proceed, we need to recall the central ideas and results from our earlier work~\cite{venkatasubramanian2015much,venkatasubramanian2017inequality}, and so we summarize them here for the benefit of the reader. 
    
    Let us first recall the expression we derived for the utility of a rational agent employed in a company. We arrived at this by seeking to answer the basic question of why people seek employment. At the most fundamental, \emph{survival}, level, it is to be able to pay bills now so that they can make a living, with the hope that the current job will lead to a better future. One hopes that the present job will lead to a better one next, acquired based on the experience from the current job, and to a series of better jobs in the future, and hence to a better life.  Thus, the utility derived from a job is made up of two components: the {\em immediate} benefits of making a living (i.e.,~``present'' utility) and the prospects of a better {\em future} life (i.e.,~``future'' utility). There is, of course, the cost or \emph{disutility} of effort or contribution to be accounted for as well. 

	Hence, we proposed that the \emph{effective} utility from a job is determined by these three dominant elements: (i) utility from salary, (ii) disutility of effort or contribution, and (iii) utility from a fair opportunity for future prospects. By effort, we do not mean just the physical effort alone, even though it is a part of it.

Thus, the effective utility for an agent is given by:
\begin{equation}
{h}_i(S_i,N_i)=u_i -v_i+w_i\label{utility_1}
\end{equation} 
where $h_i$ is the effective utility of an employee earning a salary $S_i$  by expending an appropriate  effort, while competing with ($N_i - 1$) other agents in the same job category $i$ for a fair shot at a better future. ${u}(\cdot)$ is the utility derived from salary, ${v}(\cdot)$ the disutility from effort, and ${w}(\cdot)$ is the utility from a fair opportunity for a better future.  Every agent tries to maximize its effective utility by picking an appropriate job category $i$.  

\subsection{Utility of a Fair Opportunity for a Better Future}

The first two elements are rather straightforward to appreciate, but the third requires some discussion. Consider the following scenario. A group of freshly minted law school graduates (totaling $N_i$) have just been hired by a prestigious law firm as associates.  They have been told that one of them will be promoted as partner in eight years depending on her or his performance. Let us say that the partnership is worth $\$Q$. So any associate's chance of winning the coveted partnership goes as  $1/N_i$, where $N_i$ is the number of associates in her peer group $i$, her {\em local competition}. Therefore, her expected value for the award is $Q/N_i$,  and the utility derived from it goes as $\ln (Q/N_i)$ because of diminishing marginal utility. Diminishing marginal utility is a well-known economics concept that states that the incremental benefit derived by consuming an additional unit of a resource decreases as one consumes more and more of it. This property is usually modeled by a logarithmic function. It is important to recognize here that the benefit ($\ln Q$) lies in the future, but its cost or disutility ($\ln (1/N_i)$) is paid in the present, in the daily competition with one's peers towards the partnership. This is akin to buying a lottery ticket. The cost (say, \$1) of the ticket  is incurred right away, right at the purchase, but the benefit of potentially winning a large sum lies in the future. For the time being, one is out \$1---this disutility is to be accounted for right away. 

	Therefore, the disutility incurred towards a fair opportunity for career advancement in a competitive environment is:
\begin{equation}
w_i(N_i)=-\gamma\ln N_i\label{payoff_w}
\end{equation} 
where $\gamma$ is a constant parameter. This equation models the effect of {\em competitive interaction} between agents. Considering the society at large, this equation captures the notion that in a fair society, an appropriately qualified agent with the necessary education, experience, and skills should have a fair shot at growth opportunities irrespective of her race, color, gender, and other such factors---i.e.,~it is a {\em fair competitive} environment. This is the cost or the disutility incurred for {\em equality of access} or \emph{equality of opportunity} for a better life, for {\em upward mobility}. The category $i$ would correspond to her qualification category in the society. The other agents in that category are the ones she will be competing with for these growth opportunities.    

\subsection{Modeling the Disutility of a Job}

For the utility derived from salary, we again employ the commonly used logarithmic utility function:
\begin{equation}
u_i(S_i)=\alpha\ln S_i\label{payoff_u}
\end{equation}
where $\alpha$ is another constant parameter. As for the effort component, every job has certain disutility associated with it. This disutility depends on a host of factors such as the investment in education needed to qualify oneself for the job, the experience to be acquired, working hours and schedule, quality of work, work environment, company culture, relocation anxieties, etc. To model this, one can combine $u$ and $v$ to compute
\begin{equation}
u_\text{net} = au - bv\quad\text{($a$ and $b$ are positive constant parameters)}
\end{equation} 
which is the net utility (i.e.,~net benefit or gain) derived from a job after accounting for its cost. Typically, net utility will increase as $u$ increases (because of salary increase, for example). However, generally, after a point, the cost has increased so much, due to personal sacrifices such as working overtime, missing quality time with family, giving up on hobbies, job stress resulting in poor mental and physical health, etc., $u_\text{net}$ begins to decrease after reaching a maximum. The simplest model of this commonly occurring inverted-U  profile is a quadratic function, as in
\begin{equation}
u_\text{net} = au - bu^2.
\end{equation}
Since, $u\sim\ln(\text{\em Salary})$, we get equation~\eqref{payoff_v}: 
\begin{equation}
v_i(S_i)=\beta(\ln S_i)^2\label{payoff_v}
.\end{equation}

\subsection{Effective Utility from a Job}

	Combining all three, we have 
\begin{equation}
{
{h}_i(S_i,N_i)=\alpha\ln S_i-\beta(\ln S_i)^2-\gamma\ln N_i\label{utility_1_new}}
\end{equation} 
where 
$\alpha,\beta,\gamma>0$.

	In general, $\alpha$, $\beta$ and $\gamma$, which model the relative importance an agent assigns to these three elements, can vary from agent to agent. However, we examined the ideal situation where all agents have the same preferences and hence treat these as constant parameters. 

\subsection{Equilibrium Income Distribution}

	We then used the potential game theoretic framework to prove~\cite{venkatasubramanian2015much,venkatasubramanian2017inequality} that a large population of agents with this utility function will reach \gls{ne}. In potential games~\cite {rosenthal1973class, sandholm2010population, easley2010networks}, there exists a single scalar-valued global function, called a {\em potential}, that captures  the necessary information about utilities. The {\em gradient} of the potential is the {\em utility}. For such games, \gls{ne} is reached when the potential is maximized. 
    
	So, using the potential game formalism, we have an employee's utility as the gradient of potential $\phi(\bm{x})$, i.e.,
\begin{equation}
{h}_i(\bm{x})\equiv \frac{\partial \phi(\bm{x})}{\partial x_i}
\end{equation}
where $x_i=N_i/N$ denotes the fraction of population at category $i$
and $\bm{x}$ is the population vector. Therefore, by integration (we replace partial derivative with total derivative because ${h}_i(\bm{x})$ can be reduced to ${h}_i(x_i)$ expressed in \eqref{utility_1}--\eqref{payoff_v}), 
\begin{equation}
\phi(\bm{x})=\sum_{i=1}^n\int {h}_i(\bm{x})\mathrm{d}x_i
.\end{equation}

We observe, using \eqref{utility_1_new}, that our game is a potential game with the potential function \begin{equation}
\phi(\bm{x})=\phi_{u}+\phi_v+\phi_w+\text{constant}\label{pay_potential}
\end{equation}
where
\begin{align}
\phi_u&=\alpha\sum_{i=1}^nx_i\ln S_i\\
\phi_v&=-\beta \sum_{i=1}^nx_i(\ln S_i)^2\\
\phi_w&=\frac{\gamma}{N} \ln \frac{N!}{\prod_{i=1}^n(Nx_i)!}\label{fair_potential}
\end{align}
where we have used Stirling's approximation in equation~\eqref{fair_potential}.

	One can see that $\phi(\bm{x})$ is strictly concave:
\begin{equation}
\frac{\partial^2 \phi(\bm{x})}{\partial x_i^2}=-\frac{\gamma}{x_i}<0
\end{equation}
	Therefore, {\em a unique \gls{ne} for this game exists}, where $\phi(\bm{x})$ is maximized, as per the well-known theorem in potential games~\cite[p. 60]{sandholm2010population}. 

	Thus, the self-organizing free market dynamics, where employees switch jobs, and companies switch employees, in search of better utilities or profits, ultimately reaches an equilibrium state, with an {\em equilibrium income distribution}. This happens when the potential $\phi(\bm{x})$ is maximized. The equilibrium income distribution is the following {\em lognormal distribution}:

\begin{equation}
x_i=\frac{1}{S_iD}\exp\left[-\frac{\left(\ln S_i-\frac{\alpha+\gamma}{2\beta}\right)^2}{\gamma/\beta}\right]\label{logn_potential}
\end{equation}

where $D=N\exp\left[\lambda/\gamma-(\alpha+\gamma)^2/4\beta\gamma\right]$ and $\lambda$ is the Lagrange multiplier
used in maximizing $\phi(\bm{x})$ with the $\sum_{i=1}^{n}x_{i}=1$ constraint.

	We also proved that the effective utility, $h^*$, at equilibrium is given by:
\begin{equation}
h^*=\gamma\ln Z - \gamma\ln N
\end{equation} 

where $Z=\sum_{j=1}^n\exp\big([\alpha\ln S_j-\beta(\ln S_j)^2]/\gamma\big)$ resembles the partition function seen in statistical mechanics (it is easy to show that $\lambda = h^*$). \emph{At equilibrium, all agents enjoy the same effective utility or effective welfare, $h^*$.}  Everyone is {\em not} making the same salary, of course, but they all enjoy the same effective utility. This is an important result, for it proves that all agents are treated \emph{equally} with respect to the economic outcome, namely, effective utility. This proves that the ideal free-market exhibits \emph{outcome fairness}. 

	We also proved that this distribution is socially optimal. A socially optimal distribution is one where the effective utility of the entire population (i.e.,~the {\em total effective utility} of society, $H$) is maximized. This is the outcome desired by utilitarians such as Jeremy Bentham and John Stuart Mill. This \emph{maximum} total effective utility ($H^*$) is given by    
	
\begin{equation}
H^* = \sum_{i=1}^n N_i h_i^{\ast} = \sum_{j=1}^N h_j^{\ast} = Nh^*
\end{equation}
(since $h_i^{\ast} = h_j^{\ast} = h^*$ at equilibrium), subject to the constraints 

\begin{align}
\sum_{i=1}^n N_i S_i&= M\label{m_condition}\\
\sum_{i=1}^n N_i&=N\label{n_condition}
\end{align}

Note that the index $i$ covers the $n$ different salary levels $S_i$, whereas $j$ covers the $N$ employees. $M$ is the total salary budget and $N$ is the total number of employees in a company. 

Now that we have covered the necessary background work, we are ready to forge ahead to develop the \glsentrylong{nbs} framework and its connection to maximum entropy. Towards that, let us first provide an intuitive explanation of the maximum entropy principle.

\section{Entropy as a Measure of Fairness: An Intuitive Explanation of $S = -\sum_{i=1}^n p_i\ln p_i$}

	As we know, entropy is maximized at statistical equilibrium. We also know that the equivalent defining criterion for statistical equilibrium is the {\em equality} of all accessible phase space cells, which, in fact, is the fundamental postulate of statistical mechanics. In other words, a given molecule is {\em equally likely} to be found {\em in any one} of the accessible phase space cells, at any given time. 
    
    For example, consider a very large number of identical gas molecules enclosed in a chamber. To say that a gas molecule, in the long run, is more likely to be found in the left half of the chamber than the right, and assign it a higher probability (say, $p(\text{left}) = 0.6$ and $p(\text{right}) = 0.4$), would be a {\em biased} and an {\em unfair} assignment of probabilities. This assignment {\em presumes} some information that has not been given. What does one know about the molecule, or its surrounding space, that would make one prefer the left chamber over the right? There is no basis for this preference. The unequal assignment of probabilities is thus arbitrary and unwarranted. Therefore, the {\em fairest} assignment of probabilities is one of {\em equality} of outcomes, i.e.,~$p(\text{left}) = 0.5$ and $p(\text{right}) = 0.5$. 
    
    Let us examine this example further. Let us say that the chamber is divided into $n$ imaginary partitions of equal volume such that the molecules are free to move about from partition to  partition. As noted, the essence of statistical equilibrium is the \emph{equality} of all accessible phase space cells. But how is this connected to the maximization of entropy as defined by $S = -\sum_{i=1}^n p_i\ln p_i$ (not to be confused with salary $S$)? The connection is not obvious from this equation. However, there is a fascinating connection with an important implication on the true nature of entropy, as well as for the Nash product. 
    
    Now, the criterion for equilibrium in our chamber example is 
\begin{equation}\label{eq:equal-probability}
p_1 = p_2 = \ldots = p_n = p
\end{equation}
where $p_i$ is the probability of finding a given molecule in the partition $i$. 

\emph{But how do we recognize and enforce this criterion in practice?}

	This is easy to do, for example, if there are only a few partitions in phase space, e.g., two partitions, left and right, as we did above. We would immediately recognize a non-equilibrium situation if one had $p(\text{left}) = 0.6$ and $p(\text{right}) = 0.4$. And we would also recognize with equal ease the equilibrium situation, $p(\text{left}) = 0.5$ and $p(\text{right}) = 0.5$. 

	But this gets trickier when there are a large number of partitions or cells in phase space. For example, what if there are 1000 partitions, as opposed to just two, and we are given the following two situations to evaluate?
\begin{equation*}
p_1 = 0.0015, p_2 = 0.001, p_3 = 0.0009, \ldots, p_{1000} = 0.0011
\end{equation*}
\begin{equation*}
p_1 = 0.0011, p_2 = 0.0008, p_3 = 0.0012, \ldots, p_{1000} = 0.0009
\end{equation*}
Since all the partitions are of equal volumes, at equilibrium,     
\begin{equation*}
p_1 = p_2 = ... = p_{1000} = 0.001
\end{equation*}   
	How do we compare these two configurations and determine which one is \emph{closer} to equilibrium? In general, how do we make this comparison among billions of such alternative configurations? How do we enforce this equality criterion? 
    
	One could compute the sum of squared residuals for comparison, or other such methods, but this gets to be really tedious and messy, when the number of molecules and the number of configurations run into billions as is typical in statistical mechanics. Is there a simpler, more elegant way of accomplishing this? It turns out there is, one that exploits a wonderful result regarding the product of a set of positive numbers that sum to a fixed amount. This is what is at the heart of the entropy equation, and the Nash product as we show later.

	While we recognize the equilibrium state by the maximum entropy criterion, hidden in its mathematical form is the \emph{equality} of all accessible cells criterion given by equation~\eqref{eq:equal-probability}. To see this, we first observe that maximizing \big\{$-\sum_{i=1}^n p_i\ln p_i$\big\} is the same as maximizing \big\{$-\ln \prod_{i=1}^n p_i^{p_i}$\big\}. And this is the same as minimizing \big\{$\ln \prod_{i=1}^n p_i^{p_i}$\big\}, which is the same as minimizing \big\{$\prod_{i=1}^n p_i^{p_i}$\big\}. Now, according to two well-known results (which are related), the arithmetic mean-geometric mean inequality theorem (AM\textendash GM theorem)~\cite{lohwater1982inequalities}, and the Jensen's inequality, this product is minimized if and only if 
\begin{equation*}
p_1 = p_2 = .... = p_n
\end{equation*}
 where $0 < p_i < 1$ and $\sum_{i=1}^n  p_i = 1$ (one can also demonstrate this using the Lagrangian multiplier method, but, we believe, the geometric intuition is more transparent using the AM-GM theorem or the Jensen's inequality). This equality is, of course, the fundamental criterion for statistical equilibrium. Therefore, when we maximize entropy, what we are really doing is \emph{enforcing this equality constraint} buried in it. This is a clever mathematical trick for enforcing equality, by exploiting this important relationship between the product and the sum of a set of positive numbers. Therefore, enforcing equality and hence fairness, is the objective buried in the mathematical trick of maximizing the product, \big\{-$\prod_{i=1}^n p_i^{p_i}$\big\}, which is the same as minimizing \big\{$\prod_{i=1}^n p_i^{p_i}$\big\}. We will now see in the next section how the same clever mathematical device of maximizing a product to test and enforce equality is employed for the Nash product.  

  	We can see clearly now how entropy maximization is related to enforcing equality, which is the same as enforcing \emph{fairness}---treating equal things equally. This insight that entropy really is a measure of fairness in a distribution has never been clearly recognized in statistical mechanics, in information theory, or in economics literature until the first author discussed its importance in his  2009\textendash2010 papers~\cite{venkatasubramanian2009fair,venkatasubramanian2010fairness}.    It is a historical accident that the concept of entropy was first discovered in the context of thermodynamics and, therefore, it has generally been identified as a measure of randomness or disorder. However, the true essence of entropy is fairness, which appears with different masks in different contexts (for a detailed discussion of this, see\cite{venkatasubramanian2017inequality}).

\section{Nash Bargaining Solution}\label{sec_nbs}

We now derive the \glsentrylong{nbs} to our problem. Following Nash's original formulation~\cite{nash1950bargaining} for the two-player bargaining problem there has been extensive literature on this topic, which has been generalized to the $n$-player case using both cooperative and noncooperative approaches with many applications  (we cite a few select papers here~\cite{harsanyi1963nperson, rubinstein1982bargaining, binmore1986nash, mazumdar1991fairness, chatterjee1993nash, krishna1996multilateral, muthoo1999bargaining, mazumdar2000game, ray2007coalition, okada2010nash, compte2010coalitional}). 

We consider $N$ players labeled $j = 1,\ldots,N$ {(use $j$ instead of $i$ for the index of individual players), and we consider only the grand coalition involving all $N$ players because all $N$ players are required for the company to succeed. 
The surplus that this coalition generates is non-negative. Once this team forms, the game stops. 

Suppose that players $1,\ldots,N$, have  complementary skills which can be combined to produce a divisible pie of some total utility. The pie is produced and divided among the $N$ players \emph{only} when all players
reach an agreement on their shares $(h_1, h_2,\ldots,h_N)$, where  $h_{j}$ is the share or the utility of the $j$th player. 

The following description of the \gls{nbs} is adapted from~\cite{mazumdar1991fairness}:
\begin{quote}
Consider a cooperative game of $N$ players (e.g.,~employees in a company). Let each individual player $j$ have an utility function $h_{j}(\bm{x}): \mathbb{X}\to \mathbb{R}$
where $\mathbb{X}$ is a convex, closed, and bounded set of $\mathbb{R}^{N}$. For example, in communication networks $\mathbb{X}$ denotes
the space of throughputs. Let $\bm{h}_{\text{d}} = [h_{\text{d},1},h_{\text{d},2},\ldots,h_{\text{d},N}]$ where
$h_{\text{d},j}= h_{j}(\bm{x}_{\text{d}})$ for some $\bm{x}_{\text{d}}\in\mathbb{X}$ denote a disagreement
point which all the players agree to as a starting point for
the game. In general, $\bm{h}_\text{d}$ can be thought of as the vector of
individual default utilities which the users would like to achieve, at least, if they enter the game. It is also referred to as the threat point.
Let $[\mathbb{H},\bm{h}_\text{d}]$ denote the
game defined on $\mathbb{X}$ with initial disagreement point $\bm{h}_\text{d}$ where $\mathbb{H}$
denotes the image of the set $\mathbb{X}$ under $h(\cdot)$, i.e.,~$h(\mathbb{X}) = \mathbb{H}$.
Let $F[\cdot,\bm{h}_\text{d}]:\mathbb{H}\to\mathbb{H}$ be an arbitration strategy. Then $F$ is
said to be a \gls{nbs} if it satisfies the four
axioms below.

\begin{enumerate}
	\item Let $\psi(\bm{h})=\bm{h'}$ where $h_{j}'=a_{j}h_{j}+b_{j}$ for $j=1,2,\ldots,N$ and $a_{j}>0$, $b_{j}$ 
	are arbitrary constants. Then 
	\begin{equation}
	F[\psi(\mathbb{H}),\psi(\bm{h}_\text{d})] = \psi(F[\mathbb{H},\bm{h}_\text{d}]).
	\end{equation}
	This states that the operating point in the space of strategies is \emph{invariant} with respect to 
	\emph{linear} utility transformations.
	
	\item The solution must satisfy 
	\begin{equation}
	(F[\mathbb{H},\bm{h}_\text{d}])_{j}\geq h_{\text{d},j}\quad(j=1,2,\ldots,N)
	\end{equation}
	and furthermore there exists no $\bm{h}\in \mathbb{H}$ such that $h_{j}\geq (F[\mathbb{H},\bm{h}_\text{d}])_{j}$ 
	for all $j=1,2,\ldots,N$. This is the notion of \emph{Pareto optimality} of the solution.
	
	\item Let $[\mathbb{H}_{1},\bm{h}_\text{d}]$ and $[\mathbb{H}_{2},\bm{h}_\text{d}]$ be two games with the same initial
	agreement point such that:
	\begin{align}
	\mathbb{H}_{1} &\subset \mathbb{H}_{2}\\
	F[\mathbb{H}_{2},\bm{h}_\text{d}]&\in \mathbb{H}_{1}.
	\end{align}
	Then $F[\mathbb{H}_{1},\bm{h}_\text{d}]=F[\mathbb{H}_{2},\bm{h}_\text{d}].$ 
	This is called the \emph{independence of irrelevant alternatives} axiom.
	This states that the \gls{nbs} of a game with a larger set of strategies is the
	same as that of the smaller game if the arbitration point
	is a valid point for the smaller game.
	The additional strategies are superfluous. 
	
	\item Let $\mathbb{H}$ be symmetrical with respect to a subset $\mathbb{J}\subseteq\{1,2,\ldots,N\}$
	of indices (i.e.,~let $j,k\in \mathbb{J}$ and $j<k$), then 
	\begin{equation}\{h_{1},h_{2},\ldots,h_{j-1},h_{k},h_{j+1},\ldots,h_{k-1},h_{j},h_{k+1},\ldots,h_{N}\}\in \mathbb{H}.\end{equation}
	If $h_{\text{d},{j}}=h_{\text{d},{k}}$, then $(F[\mathbb{H},\bm{h}_\text{d}])_{j}=(F[\mathbb{H},\bm{h}_\text{d}])_{k}$ for
	$j,k\in \mathbb{J}$.
	This is the axiom of \emph{symmetry} which states that if the set of utilities is symmetric then, for any two players, if the initial agreement point corresponds to equal performance, then their arbitrated values are equal.
\end{enumerate}
\end{quote}
Nash proved that the unique solution, i.e.,~the \gls{nbs}, of the game that satisfies the four axioms
is given by the point which maximizes the expression $\prod_{j=1}^{N}(h_{j}(\bm{x})-h_{\text{d},j})$, known as the \emph{Nash product}. This can be written as 
\begin{equation}\label{nbs_def}
\max\prod_{j=1}^{N}(h_{j}(\bm{x})-h_{\text{d},j})= \max\prod_{j=1}^{N}(h_j-h_{\text{d},j})
\end{equation}
where $h_{j}$ denotes both the $j$th player's utility function and its utility.

\section{\Glsentrytext{nbs} of the Income Distribution Game}

In this section, 
we define the set of feasible utilities of the income game
and show that it is a convex set.
By maximizing the product of utilities
over such set, we obtain the \gls{nbs} of the income game
and it is identical to the \gls{ne} solution we had reported earlier~\cite{venkatasubramanian2015much}.

Recall the utility function $h_{i}(S_{i},N_{i})$ defined in \eqref{utility_1_new}.
Suppose the $n$ salary levels are predetermined.
Then the utility for agents at each salary level solely
depends on its occupancy, that is
\begin{equation}\label{nbs_util}
h_{i}(N_{i}) = h_{0,i} -\gamma\ln N_{i}
\end{equation}
where $h_{0,i}\equiv \alpha \ln S_{i} - \beta (\ln S_{i})^{2}$
is a constant unique to the $i$th salary level.

We now define the set of population $\mathbb{N}$
\begin{equation}
	\mathbb{N}\equiv\{N_{i}\in\mathbb{Z}_{+}:\sum_{i=1}^{n}N_{i}\leq N\}
\end{equation}
where the total number of agents 
does not exceed $N$.
It is easy to verify that
$\mathbb{N}$ is convex.
Next, we construct the set of utilities $\mathbb{H}$
through the following mapping:
\begin{equation}
\mathbb{H}\equiv
\{h_{i} = h_{0,i}-\gamma\ln N_{i}:\sum_{i=1}^{n}N_{i}\leq N\}
.\end{equation}

To show that $\mathbb{H}$ is also convex,
let $\bm{h}_{1}=\bm{h}_{0}-\gamma\ln\bm{N}_{1}$ and 
$\bm{h}_{2}=\bm{h}_{0}-\gamma\ln\bm{N}_{2}$
be any two vectors of utilities in $\mathbb{H}$.
They corresponds to
two vectors of occupancies $\bm{N}_{1}\in\mathbb{N}$
and $\bm{N}_{2}\in\mathbb{N}$.
Define $\bm{h}_{3}$ as the 
convex combination of $\bm{h}_{1}$ and $\bm{h}_{2}$:
\begin{equation}
\begin{aligned}
\bm{h}_{3} &\equiv t\bm{h}_{1} + (1-t)\bm{h}_{2}\\
&=\bm{h}_{0}-\gamma[t\ln \bm{N}_{1} + (1-t)\ln\bm{N}_{2}]\\
&=\bm{h}_{0}-\gamma\ln\bm{N}_{3}
\end{aligned}
\end{equation}
where $0\leq t\leq 1$ is the ratio of the convex combination
and $\bm{N}_{3}\equiv e^{t\ln \bm{N}_{1} + (1-t)\ln\bm{N}_{2}}$.
We have
\begin{equation}
\begin{aligned}
\sum_{i=1}^{n}N_{i,3}&=\sum_{i=1}^{n}e^{t\ln N_{i,1} + (1-t)\ln N_{i,2}}\\
&\leq\sum_{i=1}^{n}\left[te^{\ln N_{i,1}} + (1-t)e^{\ln N_{i,2}}\right]\\
&=\sum_{i=1}^{n}\left[tN_{i,1} + (1-t)N_{i,2}\right]\\
&= t\sum_{i=1}^{n}N_{i,1} + (1-t)\sum_{i=1}^{n}N_{i,2}\\
&\leq tN + (1-t)N\\
&=N
\end{aligned}
\end{equation}
where $N_{i,j}$ is the $i$th component of the vector $\bm{N}_{j}$.
The first inequality 
\begin{equation}
e^{t\ln N_{i,1}+(1-t)\ln N_{i,2}} \leq te^{\ln N_{i,1}} + (1-t)e^{\ln N_{i,2}}
\end{equation}
follows Jensen's inequality where
\begin{equation}
f\big(tx_{1}+(1-t)x_{2}\big)\leq tf(x_{1}) + (1-t)f(x_{2})
\end{equation}
if the function $f(x)$ is convex;
and the second
follows the definition of $\mathbb{N}$.
$\bm{h}_{3}$ is therefore also in $\mathbb{H}$
because $N_{3}$ is in $\mathbb{N}$.

Recall that the \gls{nbs} is obtained by maximizing the
Nash product in \eqref{nbs_def}. 
By grouping players with the same amount of utility together 
(i.e.,~they are in the same income level), we convert the 
original \gls{nbs} problem to the following problem:

\begin{equation}\label{nash_prod_equiv}
\max_{\bm{x}\in\mathbb{X}}\prod_{j=1}^{N}(h_{j}(\bm{x})-h_{\text{d},j}) = \max_{\bm{h}\in\mathbb{H}}\prod_{i=1}^{n}h_{i}^{N_{i}}
\end{equation}
where we have set the disagreement point or the threat point of the $j$th player $h_{\text{d},j}$
to be zero (i.e.,~a player agrees to enter the game as long as her effective utility is greater than zero). Note that the first product is over all the players ($N$) whereas the second is over all the levels ($n$). 

Therefore, \eqref{nash_prod_equiv} is equivalent to solving 
\begin{equation}\label{nbs_program}
\begin{aligned}
\max \quad&g(\bm{N})=\sum_{i=1}^{n}N_{i}\ln h_{i}\\
\text{s.t.}\quad&l(\bm{N})=\sum_{i=1}^{n}N_{i}-N\leq 0
\end{aligned}
\end{equation}
because logarithm is continuous and monotonic.
The \gls{kkt} necessary conditions of \eqref{nbs_program} are as follows:
\begin{align}
\nabla g(\bm{N}^{\ast}) &= \mu \nabla l(\bm{N}^{\ast})\label{kkt}\\
l(\bm{N}^{\ast})&\leq 0\\
\mu&\geq 0\\
\mu l(\bm{N}^{\ast}) &=0\label{comp_slack}
\end{align}
where $\mu$ is a \gls{kkt} multiplier.
Expanding \eqref{kkt}, we have
\begin{align}
\ln h_{i}^{\ast} - \frac{\gamma}{h_{i}^{\ast}}&=\mu\quad(i=1,\ldots,n)\label{kkt_sol}
.\end{align}
There exists a unique solution to \eqref{kkt_sol}
where $h_{i}^{\ast} = h^{\ast}=\gamma/W(\gamma e^{-\mu})$
and $W$ denotes the Lambert-$W$ (product log) function.
Since $W(x)$ monotonically increases as $x\geq0$ increases,
$h^{\ast}$ is maximized when $\mu>0$.
From the complementary slackness in \eqref{comp_slack},
we have $l(\bm{N}^{\ast})=0$ or
\begin{equation}
\sum_{i=1}^{n}N_{i}=N\label{N_constraint}
.\end{equation}
%
%
From \eqref{nbs_util}, we have
\begin{equation}
h^{\ast} = \alpha\ln S_{i} -\beta(\ln S_{i})^{2} -\gamma \ln N_{i}^{\ast}
.\end{equation}
The \gls{nbs} of the income game therefore requires
\begin{equation}
N_{i}^{\ast} = \frac{N}{S_iD}\exp\left[-\frac{\left(\ln S_i-\frac{\alpha+\gamma}{2\beta}\right)^2}{\gamma/\beta}\right]\label{nbs_N}
\end{equation}
where $D=\exp\left[h^{\ast}/\gamma-(\alpha+\gamma)^2/4\beta\gamma\right]$. Equation 
\eqref{nbs_N} is identical to the \gls{ne} solution \eqref{logn_potential} since $\lambda = h^{\ast}$. This corresponds to all agents enjoying the same effective utility $h^*$ at equilibrium given by equation (16). 

\section{Summary and Conclusions}

	We have presented a \glsentrylong{nbs} to the problem of fair income distribution in an ideal free-market economy. As noted, this is the first time the \gls{nbs} formalism has been proposed for this problem even though the formalism itself has been well-known for about seven decades and the fair inequality question has been open for over two centuries. 
    
In addition, since the \gls{nbs} outcome results in a lognormal distribution, which we have proved in our earlier work as the fairest outcome, we see the connection between \gls{nbs} and fairness. By maximizing the Nash product, we are indeed maximizing fairness. We also showed that by maximizing entropy one was essentially maximizing fairness in the probability distribution by enforcing the equality of all accessible cells through the mathematical device of maximizing the product $-$ln$\prod_{i=1}^n p_i^{p_i}$, which is the same as minimizing $ \prod_{i=1}^n p_i^{p_i}$, by exploiting the arithmetic mean\textendash geometric mean inequality theorem or Jensen's inequality. 
    
    In a similar manner, in \gls{nbs} one again invokes the mathematical device of maximizing a product, this time the Nash product. Both techniques achieve the same desired result \textemdash the enforcement of equality. In the case of maximum entropy, we achieve the equality of the probabilities of all accessible phase space cells \textemdash  ~$p_1 = p_2 = ...=p_n$. That is, we achieve maximizing fairness at equilibrium. Similarly for \gls{nbs}, we achieve the equality of effective utilities for all agents \textemdash ~$h_1 = h_2 = ...= h_N = h^*$. That is, we achieve maximizing fairness at equilibrium. 
    
    Thus, the true economic objective of maximizing the Nash product is to treat all agents fairly subject to the Pareto optimality constraint. Since the fairness objective is buried deeply in the mathematical device of maximizing the product, just as it is buried in the formulation of maximum entropy principle, the fairness property is not obvious even when you probe it. So, naturally, people are surprised when they find the fairness outcome in the final results. That is why we have economists somewhat mystified, making  observations such as ``Nash product seems to have escaped up to now a meaningful interpretation,'' ``Although the maximization of a product of utilities is a simple mathematical operation it lacks a straightforward interpretation; we view it simply as a technical device,'' and ``The Unreasonable Fairness of Maximum Nash Welfare''~\cite{caragiannis2016unreasonable}, as quoted earlier. Thus, our work reveals the deep and surprising connection between the Nash product, entropy, and fairness. Achieving maximum fairness is the purpose of the maximum entropy principle as well as for the maximum Nash welfare function.  Enforcing equality and hence fairness, under the given constraints, is the true objective buried in the mathematical device of maximizing a product---we see this in entropy and in Nash product.

\section*{Acknowledgement}
This work is supported in part by the Center for the Management of Systemic Risk (CMSR) at Columbia University. The authors thank Resmi Suresh for her helpful comments and   assistance with the preparation of this manuscript. 

%
%
%

\bibliography{references}

\end{document}